\documentclass[journal]{IEEEtran}

\usepackage{amsfonts}
\usepackage{amssymb}
\usepackage{subfigure}
\usepackage{amsmath}
\usepackage{amsthm}
\usepackage{mathrsfs}
\usepackage[mathscr]{euscript}
\usepackage{verbatim}
\usepackage{graphicx}
\usepackage{epstopdf}
\usepackage{enumitem}
\usepackage{caption}
\usepackage{xcolor}
\usepackage{cite}

\theoremstyle{remark} \newtheorem{remark}{Remark}

\newtheorem{lemma}{Lemma}
\newtheorem{proposition}{Proposition}
\newtheorem{definition}{Definition}

\newtheorem{corollary}{Corollary}

\theoremstyle{remark} \newtheorem{theorem}{Theorem}

\newcommand{\yvec}{\mathbf{y}}

\newcommand{\mX}{\mathcal{X}}

\newcommand{\realSet}{\mathcal{R}}

\newcommand{\E}{\mathbb{E}}

\renewcommand{\vec}[1]{\mathbf{#1}}
\renewcommand{\sf}[1]{\mathsf{#1}}

\newcommand{\argmax}{\operatornamewithlimits{argmax}}

\newcommand{\changeColor}{black}
\newcommand{\ymc}[1]{{\color{black}{#1}}}
\newcommand{\AAcolor}[1]{{\color{black}{#1}}}

\begin{document}
\IEEEoverridecommandlockouts
\title{Exploiting Diversity in Molecular Timing \\ Channels via Order Statistics
\thanks{The authors are with the Department of Electrical Engineering, Stanford University, Stanford, CA, 94305 USA. This work was presented in part at the IEEE Global Communication Conference (GLOBECOME), December 2017, Singapore, \cite{murinGlobeCom17}. This research was supported in part by the NSF Center for Science of Information (CSoI) under grant CCF-0939370, and the NSERC Postdoctoral Fellowship fund PDF-471342-2015.}
}

\author{
Yonathan Murin, Nariman Farsad, Mainak Chowdhury, and Andrea Goldsmith \\
{Department of Electrical Engineering, Stanford University, USA} \\
\vspace{-.55cm}
}
\date{}


\maketitle

\begin{abstract}
	We study diversity in one-shot communication over molecular timing channels. We consider a channel model where the transmitter {\em simultaneously} releases a large number of information particles, while the information is encoded in the {\em time of release}. 
	The receiver decodes the information based on the {\em random} time of arrival of the information particles. 
	The random propagation is characterized by the general class of right-sided unimodal densities. 	
	We characterize the {\em asymptotic exponential} decrease rate of the probability of error as a function of the number of released particles, and denote this quantity as the {\em system diversity gain}. 
	Four types of detectors are considered: the maximum-likelihood (ML) detector, a linear detector, a detector that is based on the first arrival (FA) among all the transmitted particles, and a detector based on the last arrival (LA). 
	When the density characterizing the random propagation is supported over a large interval,
	we show that the simple FA detector achieves a diversity gain very close to that of the ML detector.
	On the other hand, when the density characterizing the random propagation is supported over a small interval, we show that the simple LA detector achieves a diversity gain very close to that of the ML detector. 
\end{abstract}

\vspace{-.35cm}
\section{Introduction}	

In many communication systems it is common to modulate the information bits into the amplitude or into the phase of the transmitted signal.
In this work we consider a different transmission approach in which the information is embedded in the {\em timing} of the transmissions. 
The resulting channels are commonly referred to as {\em timing channels}. 
Communication over timing channels was studied in three main contexts: communication via queues, i.e., queuing timing channels \cite{ana96, sund00-1, sund00-2, kivayash09, sellke07, Aptel17}, molecular communications, i.e., molecular timing channels, \cite{eck07, sri12, li14, rose:InscribedPart1, rose:InscribedPart2, isit16, murinGlobeCom16, farsadGlobeCom16}, and covert (secure) timing channels \cite{ulukus16, dunn09, Ghassami17}.

We study a model for molecular timing channels where information is modulated through the {\em time of release} of information particles (see \cite{far16ST} for a detailed discussion regarding applications of molecular communications). 
These information particles represent molecules in the context of molecular communications, or tokens using the terminology of \cite{rose:InscribedPart1, rose:InscribedPart2}.
We focus on a one-shot communication scenario where the transmitter {\em simultaneously} releases multiple identical information particles, and the time of release is selected out of a set with {\em finite} cardinality. The receiver's objective is to detect this time of release. 
The released particles are assumed to randomly and independently propagate to the receiver, where upon their arrival they are absorbed and removed from the environment. 
Hence, the random delay until a particle arrives at the receiver can be represented as an additive noise term. 
Our objective in this work is to characterize the {\em asymptotic  exponential} decrease rate of the probability of error, at the receiver, as a function of the number of released particles. 
We refer to this quantity as the {\em system diversity gain}. The formal definition of diversity gain is given in Section \ref{sec:ProbForm}.

Comparing the diversity gains of different detection techniques indicates which method achieves a lower probability of error when the number of particles used for communication is large, without the need for explicitly calculating the probability of error. 
Thus, such a comparison can simplify the system design.
\AAcolor{Note that in \cite{asilomar16, ITsubmission} we also considered a molecular timing channel with diversity; however, in these works we derived upper and lower bounds on the capacity of the molecular timing channel while in the current work we study the diversity gain in the probability of error for one-shot communication. 
We believe that the results derived in this paper also provide insights into achievable schemes that will result in tighter lower bounds on the capacity.}    



Since we consider a causal molecular timing channel, we focus on propagation models characterized by noise densities with support on the {\em positive} real line. 
In particular, in molecular communications, the particles propagate to the receiver following a random Brownian path. When the propagation is based solely on diffusion, the additive noise associated with random delay follows the L\'evy distribution \cite{nanoComNet, murinGlobeCom16}. When the diffusion is accompanied by a drift, this additive noise follows the inverse Gaussian (IG) distribution \cite{sri12, li14}. In the model studied in \cite{rose:InscribedPart1, rose:InscribedPart2}, the additive noise representing the propagation of the tokens follows an exponential distribution (the exponential delay can represent systems with chemical reactions that cause the particles to decay quickly \cite{guo16}). Propagation based on diffusion without a drift, when the information particles have finite life span, was studied in \cite{isit16, ITsubmission}, while \cite{Pandey17} considered diffusion-based communication when the information particles experience exponential degradation.


Motivated by the above propagation models, in this work we study the {\em general} class of propagation delays where the associated noise density is continuous, differentiable, and {\em unimodal}.\footnote{\AAcolor{When the maximum of a probability density function of a continuous distribution is at a single value (or a continuous interval), the density is referred to as unimodal (as opposed to the case of multiple maxima which is referred to as multimodal). The local maximizing values are the modes of the density.}} 
We derive expressions for the system diversity gain associated with four types of detectors: the optimal maximum likelihood (ML) detector, a linear detector based on the mean of the arrival times, a detector that is based on the first arrival (FA) among the transmitted particles \cite{murinGlobeCom16}, and a detector that is based on the last arrival (LA) among the transmitted particles. 
One of the main results presented in \cite{murinGlobeCom16} is that in the case of a L\'evy-distributed additive noise, linear detection under multiple particle release has worse performance than linear detection based on a single particle release. 
This degradation is due to the fact that the L\'evy distribution has heavy tails that render linear processing highly suboptimal. It was further shown in \cite{murinGlobeCom16} that for a small number of released particles, the probability of error achieved by the FA detector is indistinguishable from that achieved by the ML detector; thus, this detector provides a simple and attractive alternative to ML detection for a small number of released particles. 

In this work we consider the complementary setting where the number of released particles is large. We show that if the mode of the density of the noise is at zero, for example as is the case for uniform or exponential distributions, then the FA and ML detectors {\em are equivalent}. Moreover, even if the mode is larger than zero, when the density of the noise is supported over a large interval (e.g. the positive real line), the FA detector can achieve a diversity gain {\em very close} to the one achieved by the ML detector, and can {\em significantly outperform the linear detector}. 
We emphasize that this holds {\em regardless} of the tails of the noise, and contradicts the common use of linear processing, known to maximize the signal-to-noise ratio (or minimize the bit error rate) in systems with receive diversity and additive Gaussian noise (see \cite{TV:05} and \cite[Sec. IV.C.2]{sri12}).
Our results indicate that for detection of signals transmitted over molecular timing channels (when the density of the noise is supported over a large interval), the FA detector is a much better alternative to the high-complexity ML detector as compared to linear processing.

While the FA detector performs very well for noise densities supported over the positive real line (e.g. the L\'evy and IG distributions), we further show that when the density of the noise is supported over a short interval, the FA detector can be significantly outperformed by linear detection. In this case, we show that if the mode of the density of the noise is at the maximum value of the support, then the LA and ML detectors {\em are equivalent}. Moreover, even if this condition does not hold, when the density of the noise is supported over a small interval, the LA detector can achieve a diversity gain {\em very close} to the one achieved by the ML detector, and can {\em significantly outperform the linear detector}. \AAcolor{Thus, our results indicate that detection based on order statistics of the arrival times, namely, based on the FA and LA, exploits the diversity of the channel in a near-optimal manner, thereby establishing a low-complexity near-ML detection framework for one-shot communication over timing channels.} 

The rest of this paper is organized as follows.
The problem formulation is presented in Section \ref{sec:ProbForm}. 
The diversity gain of the ML and linear detectors are derived in Section \ref{sec:PDG_ML_LIN}. The diversity gain of the FA and LA detectors is derived in Section \ref{sec:ordStat}.
Analysis of the diversity gain of specific densities is provided in Section \ref{sec:numRes}, and the paper is concluded in Section~\ref{sec:conc}.

{\bf {\slshape Notation}:} 
We denote sets with calligraphic letters, e.g., $\mathcal{X}$, where $\realSet^{+}$ denotes the set of positive real numbers.
We denote RVs with upper case letters, e.g., $X$ (except $M$ and $L$ which are used to denote constants), and their realizations with lower case letters, e.g., $x$. 
An RV takes values in the set $\mX$, and we use $|\mX|$ to denote the cardinality of a finite set. 
We use $f_{Z}(z)$ to denote the probability density function (PDF) of a continuous RV $Z$ on $\realSet^{+}$ and $F_{Z}(z)$ to denote its cumulative distribution function (CDF). 
We denote vectors with boldface letters, e.g., $\yvec$, where the $k^{\text{th}}$ element of a vector $\yvec$ is denoted by $y_k$.
Finally, we use $\log (\cdot)$ to denote the natural logarithm.

\section{Problem Formulation} \label{sec:ProbForm}

\subsection{System Model} \label{subsec:sysModel}

The molecular timing channel is illustrated in Fig. \ref{fig:diffuseMolComm}. In this channel model the information is modulated on {\em the time of release of the information particles} $X$, where $Z$ denotes the {\em random} time required for the information particle to propagate from the transmitter to the receiver. 
We make the following assumptions about the system (these assumptions are consistent with those made in previous works, for instance \cite{eck07, sri12, li14, rose:InscribedPart1, ITsubmission, nanoComNet, murinSP}):

\begin{figure}[t]
	\begin{center}
		\includegraphics[width=0.8\columnwidth,keepaspectratio]{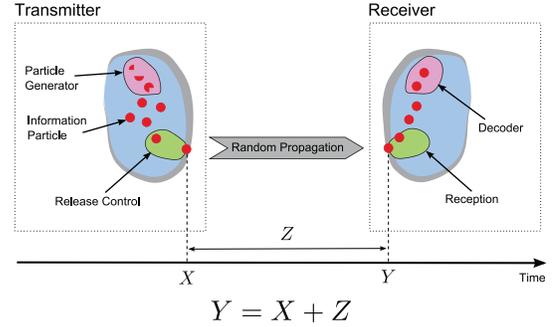}
	\end{center}
	\vspace{-0.3cm}
	\captionsetup{font=footnotesize}
	\caption{\label{fig:diffuseMolComm} Molecular communication timing channel. $X$ denotes the release time, $Z$ denote the random propagation time, and $Y$ denotes the arrival time.}
	\vspace{-0.25cm}
\end{figure}

\begin{enumerate}[label = \emph{\roman*})]
	\item \label{assmp:timeModulation}
	The information particles are assumed to be {\em identical and indistinguishable}, thus, the information is encoded {\em only} in the time of release of the particles. At the receiver, the information is decoded based {\em only} on the time of arrival. 
	
	\item \label{assmp:synch}
	The time-synchronization between the transmitter and the receiver is perfect, the transmitter perfectly controls the particles' release time, and the receiver perfectly measures their arrival time. 
	
	\item \label{assmp:Arrival}
	Every information particle that arrives at the receiver is absorbed and removed from the system.
	
	\item \label{assmp:indep}
	The information particles propagate {\em independently} of each other, and their trajectories are random according to an i.i.d. random process.
\end{enumerate} 

Let $\mX$ be a finite set of constellation points on the real line: $\mathcal{X} \triangleq \{\xi_0, \xi_1, \dots, \xi_{L-1} \}$, $0 \le \xi_0 \le \dots \le \xi_{L-1} = \Delta$. 
Observing $l \in \{0,1,\dots,L-1 \}$ with equal probability, the transmitter {\em simultaneously} releases $M$ information particles into the medium at time $X \in \mX$. The release time $X$ is assumed to be independent of the random propagation time of {\em each} of the information particles. 
Let $\{Y_m\}_{m=1}^M$ denote the $M$ arrival times of each of the information particles released at time $X$. Due to causality, we have $Y_{m} > X, m=1,2,\dots,M$. This leads to the following additive noise channel model:     
\begin{align}
	Y_{m} = {X} + Z_{m}, \quad m=1,2,\dots,M,
\label{eq:LevyChan}
\end{align}

\noindent where $Z_{m} \in \realSet^{+}$ is a random noise term representing the propagation time of the $m^{\text{th}}$ particle. Note that Assumption \ref{assmp:indep} implies that the RVs $Z_{m}$ are independent of each other. The channel model \eqref{eq:LevyChan} represents the setting of a transmitter (e.g., a nano-scale sensor) that infrequently sends a symbol (which conveys a limited number of bits) to a receiver (e.g., a centralized molecular controller), and then remains silent for a long period. Thus, the communication has a one-shot nature and there is no inter-symbol interference.
 
In this paper we restrict our attention to the case of a binary modulation, i.e., $\mX = \{ 0, \Delta \}$. We note that the derived results can be extended to more than two elements in the set $\mX$ (see the approach taken in \cite[Sec. VI]{murinSP}). It can also be extended to the case of unequal a-priori probabilities. 

Let $\hat{X}$ denote the estimation of $X$ at the receiver. We denote the probability of error, when $M$ particles are used, by $P_{\varepsilon}^{(M)} \triangleq \Pr \{X \neq \hat{X} \}$.
As all the particles are {\em simultaneously} released, $P_{\varepsilon}^{(M)}$ can decrease when $M$ is increased, \cite{sri12, murinGlobeCom16, murinSP}. Specifically, in this paper we focus on the {\em exponential} decrease of $P_{\varepsilon}^{(M)}$ in the {\em asymptotic limit of increasing} $M$, defined by the quantity $\sf{D}$ given by:
\begin{align}
	\sf{D} \triangleq \lim_{M \to \infty} \frac{-\log P_{\varepsilon}^{(M)}}{M}.
\end{align}

\begin{remark}[{\em System diversity gain}]
	The channel \eqref{eq:LevyChan} has a single input and multiple outputs. Thus, by simultaneously releasing $M$ particles we achieve {\em receive diversity}. This is the motivation for referring to $\sf{D}$ as the \ymc{\em system diversity gain}.	\textcolor{\changeColor}{Clearly, if $P_{\varepsilon}^{(M)}$ does not decrease {\em at least} exponentially with $M$, then the system diversity gain is $\sf{D} \mspace{-2mu} = \mspace{-2mu} 0$.}
	As the propagation of all particles is independent and identically distributed (see Assumption \ref{assmp:indep}), the channel \eqref{eq:LevyChan} can also be viewed as a single-input-multiple-output channel where all the channel outputs experience an independent and identical propagation law.
\end{remark}

%
We emphasize that this description of communication over a molecular timing channel is fairly general and can be applied to different propagation mechanisms as long as Assumptions \ref{assmp:synch}--\ref{assmp:indep} hold. Next, we discuss the random propagation model for our channel.

\subsection{The Random Propagation Model} \label{subsec:RandomPropag}

Our assumptions on the propagation model require the following definition of the class of weakly unimodal (quasi-concave) functions \cite[Sec. 3.4.1]{boyd-book}:

\begin{definition} \label{def:unimodal}
	A function $f(z)$ is said to be weakly unimodal if there exists a value $\zeta$ for which it is weakly monotonically increasing for $z \le \zeta$ and weakly monotonically decreasing for $z \ge \zeta$. Note that for a weakly unimodal function the maximum value can be reached for a continuous range of values of $z$.
	In the following we refer to the class of weakly unimodal functions simply as unimodal functions.
\end{definition}


We now define the proposed model associated with the timing channel \eqref{eq:LevyChan}:
\begin{definition}
	For the channel model of \eqref{eq:LevyChan}, we focus on propagation models characterized by noise $Z_m$ with a noise density function $f_Z(z)$ supported over $\realSet^{+}$.\footnote{Note that \cite{farsadGlobeCom16} considered a molecular timing channel with differential transmission where the noise density support is $\realSet$. Yet, for the channel model \eqref{eq:LevyChan}, the noise can have only positive values.} We further assume that the density function $f_Z(z)$ is {\em continuous, differentiable, and unimodal}.
\end{definition}


\begin{remark} [{\em Generality of the studied model}]
	Note that we {\em do not} restrict $f_Z(z)$ to have finite first or second moments.
\end{remark}

The above assumptions hold for the random propagation models used to characterize many molecular timing channels.
For instance, in diffusive molecular communications the released particles follow a random Brownian path from the transmitter to the receiver. In the case of diffusion {\em without} a drift, the RVs $Z_m$ are L\'evy-distributed \cite[Def. 1]{murinSP}, while in the case of diffusion {\em with} a drift, the RVs $Z_m$ follow the IG distribution \cite[eq. (3)]{sri12}. Moreover, if the information particles have a finite deterministic life time (see \cite{ITsubmission}), then the resulting densities of the RVs $Z_m$ are the clipped-L\'evy density (diffusion without a drift) and the clipped-IG distribution (diffusion with drift).
Another example for a random propagation model that satisfies these assumptions is the exponential noise that was considered in the communication model of \cite{rose:InscribedPart1}.
We emphasize that the results derived in the following sections hold for {\em any} channel model obeying \eqref{eq:LevyChan} (not only in the case of molecular timing channels), provided that the density of the noise is continuous, differential, right-sided, and unimodal.


Next, we derive the diversity gain of the ML detector and the linear (mean) detector.
	
\section{System Diversity Gain of the ML and Linear Detectors} \label{sec:PDG_ML_LIN}

\subsection{The ML Detector} \label{subsec:ML}

Let $\yvec = \{ y_m\}_{m=1}^M$. The ML detector is given by the following decision rule:
\vspace{-0.1cm}
\begin{align}
	\hat{X}_{\text{ML}}(\vec{y}) = \begin{cases} 0, & \sum_{m=1}^M{\log \frac{f_Z(y_m)}{f_Z(y_m - \Delta)}}   \ge 0  \\ \Delta, &  \text{otherwise}. \end{cases} \label{eq:MLdetector}
\end{align}

\noindent For many types of $f_Z(z)$ an explicit expression for $P_{\varepsilon,\text{ML}}^{(M)}$, the probability of error of the ML detector, is not available. 
However, as recovering $x$ from the $M$ i.i.d. realizations $\{y_m\}_{m=1}^M$ belongs to the class of binary hypothesis problems, the diversity gain is exactly the Chernoff information. This is formalized in the following proposition: 
\begin{proposition} \label{prop:ML_PDG}
	The diversity gain for the ML detector in \eqref{eq:MLdetector} is given by:
	\begin{align}
		\mspace{-8mu} \sf{D}_{\text{ML}} \mspace{-3mu} = \mspace{-1mu} - \mspace{-8mu} \min_{s: 0 \leq s \leq 1} \mspace{-3mu} \log \mspace{-2mu} \left(\int_{y=\Delta}^{\infty} \mspace{-5mu} (f_Z(y))^{s} \mspace{-3mu} \cdot \mspace{-3mu} (f_Z(y \mspace{-3mu} - \mspace{-3mu} \Delta))^{1-s} dy\right). \label{eq:ML_PDG}
	\end{align}
\end{proposition}

\begin{IEEEproof}[$\mspace{-35mu}$ Proof]
	The result follows from combining \cite[Theorem 11.9.1]{cover-book} and \cite[eq. (11.239)]{cover-book} for continuous distributions, see the detailed discussion in \cite[Sec. V.B]{murinSP}.
\end{IEEEproof}

As the ML detector minimizes the probability of error for equiprobable signaling, it also {\em maximizes} the diversity gain $\sf{D}$. At the same time, the ML detector has two main drawbacks. 
First, it is relatively complicated to compute in low-complexity devices (e.g., nano-scale sensors) due to the logarithm of the quotient of the densities. Second, the ML detector requires {\em all} the particles to arrive. In some scenarios this may require very {\em long delays}, in particular when $f_Z(z)$ has heavy tails (e.g., the L\'evy distribution). 
A detector that (partially) addresses the first drawback is the linear detector discussed next.

\subsection{The Linear Detector}

If the additive noise is Gaussian, i.e., $Z_m \sim \mathcal{N}(0,\sigma_z^2)$, then the optimal detector is linear \cite[Ch. 3.3]{TV:05}. Even when the noise is not Gaussian, detection based on a {\em linear combination} of the received signals $\{y_m\}_{m=1}^M$ can significantly improve the probability of error, as observed in \cite[Sec. IV.C.2]{sri12} for the case of additive IG noise. 
We note here that the main benefit of the linear detector is its simplicity; yet, it may also require long delays. Before formally defining the linear detector, we comment on the destructive effect linear detection can have on performance in the case of heavy tailed $f_Z(z)$.

\begin{remark}[{\em The destructive nature of linear detection for heavy-tailed noise}] \label{rem:destructiveLinear}
		In \cite[Thm. 1]{murinSP} it is shown that, for the case of L\'evy-distributed propagation, \AAcolor{a linear detector (e.g., applying ML detection based on the averaged arrival times)} increases the dispersion of the noise.\footnote{The dispersion of the noise is also known as its scale.} Thus, the probability of error of a linear detector for $M>1$ is {\em lower bounded} by the probability of error of an optimal detector for the case of $M=1$, which means that the linear detector has a diversity gain of zero.
\end{remark}

	To derive the diversity gain of the linear detector we use tools from large deviations theory \cite{dembo-book}.
	Consider detection based on $Y_{\text{LIN}} \mspace{-3mu} \triangleq \mspace{-3mu} \frac{1}{M} \sum_{m=1}^M \mspace{-3mu} Y_m$, and let
	\begin{align}
		\Lambda_Z(\rho) \triangleq \log \E_Z \left\{ e^{\rho Z} \right\} \label{eq:cummulantDef}
	\end{align}
	
	\noindent denote the cumulant generating function of $Z$. Further, define 
	\begin{align}
		\Lambda_Z^{\ast}(v) \triangleq \sup_{\lambda} \left\{ \lambda v - \Lambda_Z(\lambda) \right\}
	\end{align}
	
	\noindent to be the rate (Cram\'{e}r) function \cite[Sec. 2.2]{dembo-book}.
	The diversity gain of $\hat{X}_{\text{LIN}}(Y_{\text{LIN}})$, the ML detector based on $Y_{\text{LIN}}$, is stated in the following theorem.
	\begin{theorem}
		Let $f_Z(z)$ be a density with $\E \{Z\} \mspace{-3mu} = \mspace{-3mu} \mu \mspace{-3mu} < \mspace{-3mu} \infty$, and $\E \{(Z \mspace{-3mu} - \mspace{-3mu} \mu)^2 \} \mspace{-3mu} < \mspace{-3mu} \infty$. Further assume that $\Lambda_Z(\rho)$ is finite over some interval in $\realSet$. If there exists $\alpha \mspace{-3mu} \in \mspace{-3mu} (\max \{\Delta \mspace{-3mu} - \mspace{-3mu} \mu, 0\}, \Delta)$, such that $\Lambda_Z^{\ast}\left( \mu \mspace{-3mu} + \mspace{-3mu} \alpha \right) \mspace{-3mu} = \mspace{-3mu} \Lambda_Z^{\ast}\left( \mu \mspace{-3mu} - \mspace{-3mu} \Delta \mspace{-3mu} + \mspace{-3mu} \alpha \right)$, then: 
		\vspace{-0.1cm}
		\begin{align}
			\sf{D}_{\text{LIN}} = \Lambda_Z^{\ast}\left( \mu + \alpha \right). \label{eq:Lin_PDG} 
		\end{align}
		
		\vspace{-0.15cm}
		\noindent Otherwise, $\sf{D}_{\text{LIN}} = \infty$.
	\end{theorem}
	
\begin{IEEEproof}[$\mspace{-35mu}$ Proof]
	Let $P_{\varepsilon|0}^{(M)}$ and $P_{\varepsilon|\Delta}^{(M)}$ denote the probabilities of error given $x=0$ and $x=\Delta$, respectively. The diversity gain is now given by:
	\vspace{-0.1cm}
	\begin{align}
		\sf{D}_{\text{LIN}} = \min \left\{ \lim_{M \to \infty} \frac{-\log P_{\varepsilon|0}^{(M)} }{M}, \lim_{M \to \infty} \frac{-\log P_{\varepsilon|\Delta}^{(M)}}{M} \right\}. \label{eq:Lin_PDG_general}
	\end{align}
	
	\vspace{-0.15cm}
	\noindent From Cram\'{e}r's Theorem \cite[Thm. 2.2.3]{dembo-book} we have $\lim_{M\to \infty} \log \Pr \{Y_{\text{LIN}} > y_0 | X=x\} \mspace{-3mu} = \mspace{-3mu} \Lambda_Z^{\ast}\left( y_0 \right), y_0 \mspace{-3mu} > \mspace{-3mu} \mu$. 
	Similarly, Cram\'{e}r's Theorem states that $\lim_{M\to \infty} \log \Pr \{Y_{\text{LIN}} < y_\Delta | X=x\} \mspace{-3mu} = \mspace{-3mu} \Lambda_Z^{\ast}\left( y_\Delta \right), y_\Delta \mspace{-3mu} < \mspace{-3mu} \mu$. Recalling that $f_Z(z)$ is not necessarily symmetric, to maximize $\sf{D}_{\text{LIN}}$ we use the fact that the two densities differ only in a shift and require the two terms on the right-hand-side (RHS) of \eqref{eq:Lin_PDG_general} to be the same. 
	Thus, we find the point at which the right tail ($\Lambda_Z^{\ast}\left( \mu \mspace{-3mu} + \mspace{-3mu} \alpha \right)$) equals the left tail ($\Lambda_Z^{\ast}\left( \mu \mspace{-3mu} - \mspace{-3mu} \Delta \mspace{-3mu} + \mspace{-3mu} \alpha \right)$). 	
	This leads to the decision threshold $\mu + \alpha$ and to \eqref{eq:Lin_PDG}. If such a point does not exist, then the decision intervals do not overlap, which implies zero probability of error and $\sf{D}_{\text{LIN}} = \infty$.
\end{IEEEproof}

In the next section we discuss detection based on order statistics, and show that for some noise distributions the proposed detectors are equivalent to the optimal ML detector.
%

\section{Detection Based on Order Statistics} \label{sec:ordStat}

The detectors proposed in this section detect the transmitted symbol based on either the first or the last arrivals among the $M$ particles. Specifically, the detector waits for the first (last) particle to arrive and then applies ML detection based on this arrival. 
It is shown that for $M$ large enough this can be done by simply comparing the first (last) arrival to a threshold. 

\subsection{The FA Detector}

Let $y_{\text{FA}} \triangleq \min \left\{ y_1,y_2,\dots,y_M \right\}$. The FA detector is the ML detector based on $Y_{\text{FA}}$. Before discussing the performance of the ML detector, we note that for a fixed value of $M$, $f_{Y_{\text{FA}}|X}(y_{\text{FA}}|x)$ {\em is not necessarily unimodal}. 
The following lemma provides sufficient conditions for $f_{Y_{\text{FA}}|X}(y_{\text{FA}}|x)$ to be unimodal, for sufficiently large (yet finite) $M$.
\begin{lemma} \label{lemm:unimodalFA}
		Let $f_Z(z)$ be a unimodal density supported on the interval $(0, \tau), \tau \le \infty$, and $f'_Z(z)$ its derivative. 
		If there exists an $\epsilon > 0$ such that, for every $0 < z \le \epsilon$, the function $g(z) = \frac{f'_Z(z)}{f_Z^2(z)}$ is monotonically decreasing, then there exists a sufficiently large and finite $M_0$ for which the density of $Z_{\text{FA}} \mspace{-3mu} \triangleq \mspace{-3mu} \min\{Z_m\}_{m=1}^M$ is unimodal for $M>M_0$.
\end{lemma}

\begin{IEEEproof}[$\mspace{-35mu}$ Proof]
	The proof is provided in Appendix \ref{annex:proof_lemm_unimodalFA}.
\end{IEEEproof}

\begin{remark} [{\em Shifted support}]
	In Lemma \ref{lemm:unimodalFA} it is assumed that the support of $f_Z(z)$ is $(a,b), a = 0, b \le \infty$. The lemma can be easily extended to the case of $a > 0$.
\end{remark}

\begin{remark} [{\em Sufficient conditions for unimodality}]
	Lemma \ref{lemm:unimodalFA} provides {\em sufficient} conditions for the density of the FA, $f_{Y_{\text{FA}}|X}(y_{\text{FA}}|x)$, to be unimodal when $M$ is sufficiently large. It is possible that $f_{Y_{\text{FA}}|X}(y_{\text{FA}}|x)$ will be unimodal even if these conditions do not hold. It is also possible that $f_{Y_{\text{FA}}|X}(y_{\text{FA}}|x)$ will be unimodal for values of $M$ smaller than $M_0$. In Appendix \ref{annex:CondsLemmaFA} we show that the conditions of Lemma \ref{lemm:unimodalFA} hold for the IG and for the L\'evy densities.
\end{remark}

Next, we assume that $f_{Y_{\text{FA}}|X}(y_{\text{FA}}|x)$ is unimodal for a given finite $M$, and provide the detection rule and probability of error of the FA detector. 
\begin{proposition} \label{prop:FA_Det}
	Let $f_{Y_{\text{FA}}|X}(y_{\text{FA}}|x)$ be unimodal for a given value of $M$, and let $m_Z$ be the mode of $f_Z(z)$. \textcolor{\changeColor}{Further, let $F_Z(z)$ be the CDF of the noise.} Then, the ML detector based on $y_{\text{FA}}$ is given by:
	\begin{align}
		\hat{X}_{\text{FA}}(y_{\text{FA}}) = \begin{cases} 0, & y_{\text{FA}} < \theta_M \\ \Delta, &  y_{\text{FA}} \ge \theta_M, \end{cases}
		\label{eq:decisionRuleFA}
	\end{align}
	
	\noindent where $\theta_M$ is the solution of the following equation in $y_{\text{FA}}$ for $\Delta \mspace{-3mu} \le \mspace{-3mu} y_{\text{FA}} \mspace{-3mu} \le \mspace{-3mu} m_Z$:
	\begin{align}
		\frac{f_Z(y_{\text{FA}})}{f_Z(y_{\text{FA}}-\Delta)} = \left( \frac{1 - F_z(y_{\text{FA}} - \Delta)}{1 - F_z(y_{\text{FA}})} \right)^{M-1}. \label{eq:FA_decRuleEquation}
	\end{align}
	
	\noindent If \eqref{eq:FA_decRuleEquation} does not have a solution, then $\theta_M = \Delta$. Furthermore, the probability of error of the FA detector is given by:
	\begin{align}
		\mspace{-8mu} P_{\varepsilon, \text{FA}}^{(M)} & \mspace{-3mu} = \mspace{-3mu} 0.5 \left( \left( 1 \mspace{-3mu} - \mspace{-3mu} F_z(\theta_M) \right)^M \mspace{-3mu}  + \mspace{-3mu} 1 \mspace{-3mu} - \mspace{-3mu} \left(1 \mspace{-3mu} - \mspace{-3mu} F_z(\theta_M \mspace{-3mu} - \mspace{-3mu} \Delta) \right)^M \right).
		\label{eq:errProbSymbBySymbFA}
	\end{align}

\end{proposition}

\begin{IEEEproof}[$\mspace{-35mu}$ Proof Outline]
	As $f_{Y_{\text{FA}}|X}(y_{\text{FA}}|x)$ is a PDF, the ML detector amounts to comparing $y_{\text{FA}}$ to a threshold, which can be found by equating the two densities. Since the two densities differ only by a shift, if \eqref{eq:FA_decRuleEquation} does not have a solution, then $\theta_M = \Delta$. Finally, \eqref{eq:FA_decRuleEquation}--\eqref{eq:errProbSymbBySymbFA} are obtained by noting that $F_{Y_{\text{FA}}|X}(y_{\text{FA}}|x) = 1 - \left(1 - F_{Z}(y-x) \right)^M$ and $f_{Y_{\text{FA}}|X}(y_{\text{FA}}|x) = M \mspace{-3mu} \cdot \mspace{-3mu} f_{Z}(y-x) \mspace{-3mu} \cdot \mspace{-3mu} \left(1 - F_{Z}(y-x) \right)^{M-1}$.	
\end{IEEEproof}

The following theorem presents the diversity gain of the FA detector.
\begin{theorem} \label{thm:FA_PDG}
	Let $f_{Y_{\text{FA}}|X}(y_{\text{FA}}|x)$ be unimodal for all $M > M_0$. Then, the diversity gain of the FA detector is given by:
	\vspace{-0.1cm}
	\begin{align}
			\sf{D}_{\text{FA}} = - \log \left(1 - F_Z(\Delta) \right). \label{eq:FA_PDG}
	\end{align}
\end{theorem}

\vspace{-0.05cm}
\begin{IEEEproof}[$\mspace{-35mu}$ Proof]
	Before proving \eqref{eq:FA_PDG} we note that if $f_{Y_{\text{FA}}|X}(y_{\text{FA}}|x)$ is unimodal for all $M > M_0$, then $\theta_M \mspace{-3mu} \to \mspace{-3mu} \Delta$ when $M \mspace{-3mu} \to \mspace{-3mu} \infty$. 
	This follows from the extreme value theorem \cite[Thm. 1.8.4]{extremvalue-book}, which implies that the limiting distribution of the considered densities concentrates towards the release time $x$ (namely, a Dirac delta at $x$), thus $\theta_M \mspace{-3mu} \to \mspace{-3mu} \Delta$.
	
	Next, we recall that $Z_{\text{FA}} \mspace{-3mu} = \mspace{-3mu} \min\{Z_m\}_{m=1}^M$, and let $\theta_M \mspace{-3mu} = \mspace{-3mu} \Delta \mspace{-3mu} + \delta_M, \delta_M \mspace{-3mu} \to \mspace{-3mu} 0$.
	Note that $P_{\varepsilon, \text{FA}}^{(M)}$ can be bounded as follows:
	\vspace{-0.1cm}
	\begin{align}
		\frac{1}{2}\Pr \{ Z_{\text{FA}} \ge \Delta + \delta_M \}  \le P_{\varepsilon, \text{FA}}^{(M)} \le \frac{1}{2} \Pr \{ Z_{\text{FA}} \ge \Delta \}. \label{eq:PeFA_bounds}
	\end{align}
	
	\vspace{-0.05cm}
	\noindent For the RHS of \eqref{eq:PeFA_bounds}, recalling that $\Pr \{ Z_{\text{FA}} \mspace{-3mu} \ge \mspace{-3mu} \Delta\} \mspace{-3mu} = \mspace{-3mu}  (1 \mspace{-3mu} - \mspace{-3mu} F_Z(\Delta))^M$, we write
	\begin{align}
		\lim_{M \to \infty} \frac{- \log \Pr \{ Z_{\text{FA}} \ge \Delta \}}{M}  \mspace{-3mu} = \mspace{-3mu} - \log (1 \mspace{-3mu} - \mspace{-3mu} F_Z(\Delta) ).
	\end{align}
	
	\vspace{-0.05cm}
	\noindent For the left-hand-side (LHS) we have $\Pr \{ Z_{\text{FA}} \mspace{-2mu} \ge \mspace{-2mu} \Delta \mspace{-2mu} + \mspace{-2mu} \delta_M\} \mspace{-3mu} = \mspace{-3mu}  (1 \mspace{-3mu} - \mspace{-3mu} F_Z(\Delta + \delta_M))^M$. Using a Taylor expansion of $\log (1 \mspace{-3mu} - \mspace{-3mu} F_Z(\Delta + \delta_M))^M$ around $\Delta$, we obtain:
	\vspace{-0.1cm}
	\begin{align}
		& \log (1 \mspace{-3mu} - \mspace{-3mu} F_Z(\Delta + \delta_M))^M \nonumber \\
		& \qquad \mspace{-3mu} = \mspace{-3mu} M\left(\log(1 \mspace{-3mu} - \mspace{-3mu} F_Z(\Delta)) \mspace{-3mu} - \mspace{-3mu} \frac{f_Z(\Delta) \delta_M}{1 - F_Z(\Delta)} \mspace{-3mu} + \mspace{-3mu} \mathcal{O}(\delta_M^2)  \right). \label{eq:FA_Taylor} 
	\end{align}
	
	\vspace{-0.05cm}
	\noindent Therefore, since $\delta_M \mspace{-3mu} \to \mspace{-3mu} 0$, we have:
	\vspace{-0.1cm}
	\begin{align}
		\lim_{M \to \infty} \frac{- \log \Pr \{ Z_{\text{FA}} \ge \Delta + \delta_M \}}{M}  \mspace{-3mu} = \mspace{-3mu} - \log (1 \mspace{-3mu} - \mspace{-3mu} F_Z(\Delta) ). \label{eq:PDG_FA_rightDist}
	\end{align}
	
	\noindent Combining \eqref{eq:PeFA_bounds}--\eqref{eq:PDG_FA_rightDist} we conclude the proof.
\end{IEEEproof}
%

We now consider the special case where the mode of $f_Z(z)$ is zero, e.g., the uniform or exponential densities:
\begin{theorem} \label{thm:zreomode}
	Let $f_Z(z)$ be a continuous, differentiable, and unimodal density with mode $m_Z \mspace{-3mu} \ge \mspace{-3mu} 0$ and $f_Z(z) \mspace{-3mu} = \mspace{-3mu} 0, z \mspace{-3mu} < \mspace{-3mu} m_Z$. Then, the FA and ML detectors are equivalent, namely, they have the same probability of error. 
\end{theorem}

\begin{IEEEproof}[$\mspace{-35mu}$ Proof]
	We focus on the case of $m_Z \mspace{-3mu} = \mspace{-3mu} 0$. The proof for $m_Z \mspace{-3mu} > \mspace{-3mu} 0$ follows along similar lines.
	Since $m_Z = 0$, then $f_Z(y) \mspace{-3mu} \le \mspace{-3mu} f_Z(y \mspace{-3mu} - \mspace{-3mu} \Delta), y \mspace{-3mu} \ge \mspace{-3mu} \Delta$. 
	The ML detection rule can be written as:
	\begin{align}
	\hat{X}_{\text{ML}}(\vec{y}) = \argmax_{x} \prod_{m=1}^{M} f_Z(y_m|X=x). \label{eq:ML_zeroMode}
	\end{align}
	
	\vspace{-0.15cm}	
	\noindent Therefore, the ML detector declares $\hat{X}_{\text{ML}}(\vec{y}) \mspace{-4mu} = \mspace{-4mu} 0$ only if there exists $Y_m \mspace{-3mu} < \mspace{-3mu} \Delta$ (otherwise it declares $\hat{X}_{\text{ML}}(\vec{y}) \mspace{-4mu} = \mspace{-4mu} \Delta$). Since testing if there exists $Y_m \mspace{-3mu} < \mspace{-3mu} \Delta$ can be implemented based on $y_{\mathrm{FA}}$, we conclude that in this case the ML detector reduces to the FA detector, and therefore the detectors are equivalent.
	\end{IEEEproof}
	
	\smallskip
	\noindent The following corollary is a direct consequence of Thm.~\ref{thm:zreomode}. 
	
	\begin{corollary} \label{cor:zreomode}
		Under the conditions of Thm. \ref{thm:zreomode} $\sf{D}_{\mathrm{FA}} \mspace{-2mu} = \mspace{-2mu} \sf{D}_{\mathrm{ML}}$.
	\end{corollary}

As stated in Thm.~\ref{thm:zreomode}, and as exemplified in Section \ref{sec:numRes}, in some cases $\sf{D}_{\mathrm{FA}}$ is very close to the optimal diversity gain $\sf{D}_{\mathrm{ML}}$. Unfortunately, as is also shown in Section \ref{sec:numRes}, there are cases where there is a substantial gap between $\sf{D}_{\mathrm{FA}}$ and $\sf{D}_{\mathrm{ML}}$. This motivates discussing a detection based on the complement order statistics, namely the last arrival.

\subsection{The LA Detector}

Let $y_{\text{LA}} \triangleq \max \left\{ y_1,y_2,\dots,y_M \right\}$. Similarly to the FA detector, the LA detector is the ML detector based on $Y_{\text{LA}}$. Before discussing the performance of the ML detector, we note that for a given $M$, $f_{Y_{\text{LA}}|X}(y_{\text{LA}}|x)$ is not necessarily unimodal. 
The following lemma provides sufficient conditions for $f_{Y_{\text{LA}}|X}(y_{\text{LA}}|x)$ to be unimodal, for sufficiently large (yet finite) $M$.
\begin{lemma} \label{lemm:unimodalLA}
		Let $f_Z(z)$ be a unimodal density supported on $(0,\tau), \tau \le \infty$, and $f'_Z(z)$ its derivative. 
		If there exists an $\epsilon > 0$ such that, for every $\tau-\epsilon < z \le \tau$, the function $g(z) = \frac{f'_Z(z)}{f_Z^2(z)}$ is monotonically decreasing, then there exists a sufficiently large and finite $M_0$ for which the density of $Z_{\text{LA}} \mspace{-3mu} \triangleq \mspace{-3mu} \max\{Z_m\}_{m=1}^M$ is unimodal for $M>M_0$.
\end{lemma}

\begin{IEEEproof}[$\mspace{-35mu}$ Proof]
	The proof is provided in Appendix \ref{annex:proof_lemm_unimodalLA}.
\end{IEEEproof}

Next, we assume that $f_{Y_{\text{LA}}|X}(y_{\text{LA}}|x)$ is unimodal for a given finite $M$, and provide the detection rule and probability of error of the LA detector. 
\begin{proposition} \label{prop:LA_Det}
	Let $f_Z(z)$ be a unimodal density supported on $(0,\tau), \tau \le \infty$, $f_{Y_{\text{LA}}|X}(y_{\text{LA}}|x)$ be unimodal for a given value of $M$, and $m_Z$ be the mode of $f_Z(z)$. Further, let $F_Z(z)$ be the CDF of the noise $Z_m$ for all $m$. Then the ML detector based on $y_{\text{LA}}$ is given by:
	\begin{align}
		\hat{X}_{\text{LA}}(y_{\text{LA}}) = \begin{cases} 0, & y_{\text{LA}} < \vartheta_M \\ \Delta, &  y_{\text{FA}} \ge \vartheta_M, \end{cases}
		\label{eq:decisionRuleLA}
	\end{align}

	\noindent where $\vartheta_M$ is the solution of the following equation in $y_{\text{LA}}$ for $\Delta \mspace{-3mu} \le \mspace{-3mu} y_{\text{LA}} \mspace{-3mu} \le \mspace{-3mu} \tau$:
	\begin{align}
		\frac{f_Z(y_{\text{LA}})}{f_Z(y_{\text{LA}}-\Delta)} = \left( \frac{F_z(y_{\text{LA}} - \Delta)}{F_z(y_{\text{LA}})} \right)^{M-1}. \label{eq:FA_decRuleEquationLA}
	\end{align}
	
	\noindent If \eqref{eq:FA_decRuleEquationLA} does not have a solution, then $\vartheta_M = \tau$. Furthermore, the probability of error of the LA detector is given by:
	\begin{align}
		\mspace{-8mu} P_{\varepsilon, \text{LA}}^{(M)} & \mspace{-3mu} = \mspace{-3mu} \frac{1}{2} \left( F_z(\vartheta_M - \Delta)^M \mspace{-3mu}  + \mspace{-3mu} 1 \mspace{-3mu} - \mspace{-3mu} F_z(\vartheta_M)^M \right).
		\label{eq:errProbSymbBySymbLA}
	\end{align}

\end{proposition}

\begin{IEEEproof}[$\mspace{-35mu}$ Proof]
The proof follows along the same lines as the proof of Proposition \ref{prop:FA_Det}, and thus it is omitted.
\end{IEEEproof}

The following theorem presents the diversity gain of the LA detector.
\begin{theorem} \label{thm:LA_PDG}
	Let $f_{Y_{\text{LA}}|X}(y_{\text{LA}}|x)$ be unimodal for all $M > M_0$. Then, the diversity gain of the LA detector is given by:
	\vspace{-0.1cm}
	\begin{align}
			\sf{D}_{\text{LA}} = - \log \left(F_Z(\tau - \Delta) \right). \label{eq:LA_PDG}
	\end{align}
\end{theorem}

\vspace{-0.05cm}
\begin{IEEEproof}[$\mspace{-35mu}$ Proof]
	Similarly to the proof of Thm. \ref{thm:FA_PDG}, we note that if $f_{Y_{\text{LA}}|X}(y_{\text{LA}}|x)$ is unimodal for all $M > M_0$, then $\vartheta_M \mspace{-3mu} \to \mspace{-3mu} \tau$ when $M \mspace{-3mu} \to \mspace{-3mu} \infty$. 
	This follows from the extreme value theorem \cite[Thm. 1.8.4]{extremvalue-book}, which implies that the limiting distribution of the considered densities concentrates towards it maximal value $\tau$, thus $\theta_M \mspace{-3mu} \to \mspace{-3mu} \tau$.
		
	Next, we recall that $Z_{\text{LA}} \mspace{-3mu} = \mspace{-3mu} \max\{Z_m\}_{m=1}^M$, and write $\vartheta_M \mspace{-3mu} = \mspace{-3mu} \tau \mspace{-3mu} - \eta_M, \eta_M \mspace{-3mu} \to \mspace{-3mu} 0$.
	Similarly to \eqref{eq:PeFA_bounds}, $P_{\varepsilon, \text{LA}}^{(M)}$ can be bounded as follows:
	\vspace{-0.1cm}
	\begin{align}
		\Pr \{ Z_{\text{LA}} \le \tau \mspace{-3mu} - \mspace{-3mu} \eta_M \mspace{-3mu} - \mspace{-3mu} \Delta \} \mspace{-1mu}  \le \mspace{-1mu} 2 P_{\varepsilon, \text{LA}}^{(M)} \mspace{-1mu} \le \mspace{-1mu} \Pr \{ Z_{\text{LA}} \le \tau \mspace{-3mu} - \mspace{-3mu}\Delta \}. \label{eq:PeLA_bounds}
	\end{align}
	
	\vspace{-0.05cm}
	\noindent For the RHS of \eqref{eq:PeLA_bounds}, recalling that $\Pr \{ Z_{\text{FA}} \mspace{-3mu} \le \mspace{-3mu} \tau \mspace{-3mu} - \mspace{-3mu} \Delta\} \mspace{-3mu} = \mspace{-3mu}  F_Z(\tau \mspace{-3mu} - \mspace{-3mu} \Delta)^M$, we write:
	\begin{align}
		\lim_{M \to \infty} \frac{- \log \Pr \{ Z_{\text{LA}} \le \tau \mspace{-3mu} - \mspace{-3mu} \Delta \}}{M}  \mspace{-3mu} = \mspace{-3mu} - \log (F_Z(\tau \mspace{-3mu} - \mspace{-3mu} \Delta) ).
	\end{align}
	
	\vspace{-0.05cm}
	\noindent For the LHS we have $\Pr \{ Z_{\text{LA}} \mspace{-2mu} \le \mspace{-2mu} \tau \mspace{-2mu} - \eta_M \mspace{-2mu} - \mspace{-2mu} \Delta \} \mspace{-3mu} = \mspace{-3mu}  F_Z(\tau - \mspace{-2mu} \eta_M \mspace{-2mu} - \mspace{-2mu} \Delta)^M$. Similarly to \eqref{eq:FA_Taylor}, using the Taylor expansion of $\log F_Z(\tau - \mspace{-2mu} - \eta_M \mspace{-2mu} - \mspace{-2mu} \Delta)^M$ around $\tau \mspace{-3mu} - \mspace{-3mu} \Delta$, we obtain:
	
	\begin{align}
		\lim_{M \to \infty} \frac{- \log \Pr \{ Z_{\text{LA}} \le \tau \mspace{-2mu} - \mspace{-2mu} \eta_M \mspace{-2mu} - \mspace{-2mu} \Delta \}}{M}  \mspace{-3mu} = \mspace{-3mu} - \log (F_Z(\tau \mspace{-3mu} - \mspace{-3mu} \Delta) ). \label{eq:PDG_LA_rightDist}
	\end{align}
	
	\noindent Combining \eqref{eq:PeLA_bounds}--\eqref{eq:PDG_LA_rightDist} concludes the proof.
\end{IEEEproof}

We now consider the special case where the mode of $f_Z(z)$ is at $\tau$:
\begin{theorem} \label{thm:taumode}
	Let $f_Z(z)$ be a continuous, differentiable, and unimodal density with support $(0,\tau), \tau \le \infty$ and mode $m_Z \mspace{-3mu} = \mspace{-3mu} \tau$. Then, the LA and ML detectors are equivalent, namely, they have the same probability of error. 
\end{theorem}

\begin{IEEEproof}[$\mspace{-35mu}$ Proof]
	Since $m_Z = \tau$, then $f_Z(y) \mspace{-3mu} \ge \mspace{-3mu} f_Z(y \mspace{-3mu} - \mspace{-3mu} \Delta), y \mspace{-3mu} \le \mspace{-3mu} \tau$. 
	The ML detection rule can be written as in \eqref{eq:ML_zeroMode}. Therefore, the ML detector declares $\hat{X}_{\text{ML}}(\vec{y}) \mspace{-3mu} = \mspace{-3mu} 0$ only if there exists $Y_m \mspace{-3mu} > \mspace{-3mu} \tau$ (otherwise it declares $\hat{X}_{\text{ML}}(\vec{y}) \mspace{-3mu} = \mspace{-3mu} 0$). Since testing if there exists $Y_m \mspace{-3mu} > \mspace{-3mu} \tau$ can be implemented based on $y_{\mathrm{LA}}$, we conclude that in this case the ML detector reduces to the LA detector, and therefore the detectors are equivalent.
	\end{IEEEproof}
	
	\smallskip
	\noindent The following corollary is a direct consequence of Thm.~\ref{thm:taumode}. 
	
	\begin{corollary} \label{cor:taumode}
		Under the conditions of Thm. \ref{thm:taumode} $\sf{D}_{\mathrm{LA}} \mspace{-2mu} = \mspace{-2mu} \sf{D}_{\mathrm{ML}}$.
	\end{corollary}

\subsection{Comparing the FA and LA Detectors} \label{subsec:compare_FALA}

Thm. \ref{thm:zreomode} and Thm. \ref{thm:taumode} imply that the FA and the LA detectors are optimal for different types of noise densities. While the FA detector is optimal when the mode is at zero, the LA detector is optimal when the mode is at $\tau$. Thus, FA is optimal for monotonically decreasing densities, while LA is optimal for monotonically increasing densities. 
It can further be observed that for a finite $\Delta$, if the support of $f_Z(z)$ is $\realSet^+$, then $\sf{D}_{\text{LA}} = 0$. 
Therefore, the LA detector better be used only when $\tau$ is finite and not too big (see the numerical results reported in Section \ref{sec:numRes}).

\AAcolor{Next we compare the system diversity gains achieved by the two detectors as a function of $\tau$. Let $\tilde{f}_Z(z)$ be a continuous, differentiable, and unimodal density with support $\realSet^{+}$. Further, let the noise density $f_Z(z)$ be generated from $\tilde{f}_Z(z)$ via:
\begin{align}
	f_Z(z) = \begin{cases} \frac{\tilde{f}_Z(z)}{\tilde{F}_Z(\tau)}, & z \le \tau \\
												0, & \text{otherwise}, \end{cases}
\end{align}

\noindent for $0 < \tau \le \infty$. Thus, $f_Z(z)$ is a clipped version of $\tilde{f}_Z(z)$ at $\tau$, and the CDF of $Z$ is given by $F_Z(z) = \frac{\tilde{F}_Z(z)}{\tilde{F}_Z(\tau)}$. We now ask: {\em What is the range of $\tau$ where $\sf{D}_{\mathrm{FA}} \ge \sf{D}_{\mathrm{LA}}$}?}
To answer this question we compare \eqref{eq:FA_PDG} and \eqref{eq:LA_PDG} and write:
\begin{align}
		& \log \left(1 - F_Z(\Delta) \right) = \log \left(F_Z(\tau - \Delta) \right) \nonumber \\
		& \Leftrightarrow \log \left(1 - \frac{\tilde{F}_Z(\Delta)}{\tilde{F}_Z(\tau)} \right) = \log \left(\frac{\tilde{F}_Z(\tau - \Delta)}{\tilde{F}_Z(\tau)} \right) \nonumber \\
		& \Leftrightarrow \tilde{F}_Z(\tau) - \tilde{F}_Z(\Delta) = \tilde{F}_Z(\tau - \Delta). \label{eq:FA_eq_LA}
\end{align}

\noindent Let $\tau^{\ast}$ be the solution of \eqref{eq:FA_eq_LA}. Then, for a fixed $\Delta$, for $\tau \le \tau^{\ast}$ the LA detector achieves a higher system diversity gain, while for for $\tau \ge \tau^{\ast}$ the FA detector achieves a higher system diversity gain. 


We conclude this section with two remarks discussing extensions to the considered FA and LA detectors.

\begin{remark}[{\em Other order statistics}]
	While the ML detector \eqref{eq:MLdetector} is optimal for detection based on {\em all} the particle arrivals, the FA and LA detectors, namely, \eqref{eq:decisionRuleFA} and \eqref{eq:decisionRuleLA}, are optimal for detection based {\em only on the FA (LA)} of a particle. 
	One can use order statistics theory to design optimal detectors based on the first (last) $M_0 \mspace{-3mu} \le \mspace{-3mu} M$ particle arrivals. Yet, the analysis of such detectors is significantly more involved. Moreover, as indicated in the next section, combining the FA and LA detectors based on $\tau^{\ast}$ can achieve system diversity gains very close to those achieved by the ML detector.
\end{remark}

\begin{remark}[{\em Larger constellations}]
	\AAcolor{The FA, LA, linear, and ML detectors can be extended to the case of larger constellations, i.e., $L>2$. 
	In this case the ML detector requires comparing all $L$ hypotheses. On the other hand, as discussed in \cite[Sec. VI]{murinSP}, given a simple choice of the constellation points $\{\xi_l\}_{l=0}^{L-1}$ (see \cite[Fig. 4]{murinSP}), optimal detection based on $Y_{\text{FA}}$ (or $Y_{\text{LA}}$) can be implemented by comparing {\em only two} hypothesizes. These two hypothesizes can be easily found based on their modes.}
	Since for the FA detector the conditional density concentrates towards $x$ (or for the LA detector towards $\tau$), for a fixed $L$ one can find large enough $M$ such that any non-zero probability of error can be achieved. This enables sending short messages of several bits using a large number of particles. 
	Note that when $L$ scales with $M$ a more involved analysis is required. This analysis involves the rate of convergence of $\delta_M$ to zero (or $\eta_M$ to $\tau$) for the specific noise density.
\end{remark}

In the next section we explicitly evaluate the formulas derived above and the resulting diversity gains for several specific propagation densities: the uniform, exponential, IG, and L\'evy distributions. We also provide numerical analysis of the system diversity gains for the clipped L\'evy and IG distributions, as a function of the clipping parameter $\tau$.

\vspace{-.15cm}
\section{Numerical Results} \label{sec:numRes}

We begin our numerical study considering densities with support $\realSet^+$ (i.e., the case of $\tau = \infty$).

\vspace{-.15cm}
\subsection{Densities With $\tau = \infty$}

Before discussing specific densities, we recall \eqref{eq:LA_PDG} which implies that, for a finite $\Delta$, if $\tau=\infty$ then $\sf{D}_{\mathrm{LA}} = 0$. Therefore, in this subsection we focus on the diversity gain of the ML, linear, and FA detector, and do not discuss the LA detector.

\subsubsection{The Exponential Distribution}

Let $Z \sim \mspace{-3mu} \mathscr{EXP}(b)$, i.e., the exponential density with rate parameter $b \mspace{-3mu} > \mspace{-3mu} 0$.
An explicit evaluation of \eqref{eq:ML_PDG} results in $\sf{D}_{\text{ML}} \mspace{-3mu} = \mspace{-3mu} b\Delta$, and following Corollary \ref{cor:zreomode}, we have $\sf{D}_{\text{ML}} = \sf{D}_{\text{FA}}$. 
Considering the linear detector, for the exponential density $\Lambda_Z^{\ast}(v) \mspace{-3mu} = \mspace{-3mu} bv \mspace{-2mu} - \mspace{-2mu} 1 \mspace{-2mu} - \mspace{-2mu} \log (bv), v > 0$.
Moreover, it can be shown that the $\alpha$ in \eqref{eq:Lin_PDG} is given by:
\begin{align}
	\alpha = \frac{1 - e^{b \Delta} (1 - b \Delta)}{(e^{b \Delta} - 1) b}.
\end{align}

\noindent Plugging this value into \eqref{eq:Lin_PDG} results in:
\vspace{-0.1cm}
\begin{align}
	\sf{D}_{\text{LIN}} = \frac{1+e^{b \Delta}(b \Delta-1) - (e^{b \Delta}-1) \log\left(\frac{b \Delta e^{b \Delta}}{e^{b \Delta}-1}\right)}{e^{b \Delta}-1}.
\end{align} 

As an example, let $b \mspace{-3mu} = \mspace{-3mu} 1$, and consider $\Delta \mspace{-3mu} \in \mspace{-3mu} \{ 0.5, 1.5, 2.5 \}$. Table \ref{tab:exp} details the resulting diversity gains. 
\begin{table}[h]
		\begin{center}
		\footnotesize
		\begin{tabular}[t]{|c|c|c|c|}
			\hline
			  & $\Delta=0.5$  & $\Delta=1.5$ & $\Delta=2.5$ \\
			\hline
			\hline
			$\sf{D}_{\text{ML}}, \sf{D}_{\text{FA}}$  & 0.5  & 1.5  & 2.5  \\
			\hline
			$ \sf{D}_{\text{LIN}}$  & 0.0312  & 0.2729  & 0.7216 \\
			\hline 
		\end{tabular}
		\captionsetup{font=footnotesize}
		\caption{$\sf{D}_{\text{ML}}, \sf{D}_{\text{FA}}$ and $\sf{D}_{\text{LIN}}$ for $\mathscr{EXP}(1)$. \label{tab:exp}}
		\vspace{-0.25cm}
	\end{center}
\end{table}

\subsubsection{The Inverse-Gaussian Distribution}

Let $Z \mspace{-3mu} \sim \mspace{-3mu} \mathscr{IG}(\mu,b)$, i.e., the IG density with mean $\mu$ and shape parameter $b \mspace{-3mu} > \mspace{-3mu} 0$. In Appendix \ref{annex:CondsLemmaFA} it is shown that for $f_Z(z) \mspace{-3mu} \sim \mspace{-3mu} \mathscr{IG}(\mu,b)$, the conditions of Lemma \ref{lemm:unimodalFA} hold and therefore $f_Z(z)$ is unimodal for sufficiently large $M$. 
While explicitly evaluating $\sf{D}_{\text{ML}}$ seems intractable, it can be evaluated numerically. For the FA detector we use the CDF of the IG density to obtain:
\begin{align}
	\sf{D}_{\text{FA}} \mspace{-3mu} & = \mspace{-3mu} -\log\left(1 \mspace{-3mu} - \mspace{-3mu} \Phi\left(\sqrt{\frac{b}{\Delta}} \left(\frac{\Delta}{\mu} \mspace{-3mu} - \mspace{-3mu} 1 \right) \right) \right. \nonumber \\
		& \mspace{90mu} \left. - \mspace{-1mu} e^{\frac{2b}{\mu}} \Phi \left(- \mspace{-3mu} \sqrt{\frac{b}{\Delta}} \left(\frac{\Delta}{\mu} \mspace{-3mu} + \mspace{-3mu} 1 \right) \right)  \right),
\end{align}

\noindent where $\Phi(x)$ is the CDF of a standard Gaussian RV. Finally, for the IG density $\Lambda_Z(\tau) \mspace{-3mu} = \mspace{-3mu} \frac{b}{\mu}\left(1 \mspace{-3mu}- \mspace{-3mu}\sqrt{1 \mspace{-3mu} - \mspace{-3mu} \frac{2\mu^2 \tau}{b}} \right), \tau \mspace{-3mu} \le \mspace{-3mu} \frac{b}{2\mu^2}$. Explicitly calculating $\Lambda_Z^{\ast}(v)$ for the IG density we obtain:
\begin{align}
	\Lambda_Z^{\ast}(v) \mspace{-3mu} = \mspace{-3mu} \frac{b \left(-\mu^2 + 2 \mu \left(\frac{\mu}{v} -1 \right)v + v^2 \right)}{2 \mu^2 v}. \label{eq:IG_RateFunc}
\end{align}

\noindent While for the IG density finding an explicit expression for $\alpha$ seems intractable, it can be found numerically using \eqref{eq:IG_RateFunc}. Table \ref{tab:IG} details the diversity gain for $\mu \mspace{-3mu} = \mspace{-3mu} 1, b \mspace{-3mu} = \mspace{-3mu} 1$, and $\Delta \mspace{-3mu} \in \mspace{-3mu} \{0.5, 1, 1.5 \}$. 
\begin{table}[h]
		\begin{center}
		\footnotesize
		\begin{tabular}[t]{|c|c|c|c|}
			\hline
			  & $\Delta=0.5$  & $\Delta=1$ & $\Delta=1.5$ \\
			\hline
			\hline
			$\sf{D}_{\text{ML}}$  & 0.4766  & 1.1070  &  1.6657 \\
			\hline
			$\sf{D}_{\text{FA}}$  & 0.4541  & 1.1029  & 1.6648  \\
			\hline
			$ \sf{D}_{\text{LIN}}$  & 0.0308  & 0.1180  & 0.2499  \\
			\hline 
		\end{tabular}
		\captionsetup{font=footnotesize}
		\caption{$\sf{D}_{\text{ML}}, \sf{D}_{\text{FA}}$ and $\sf{D}_{\text{LIN}}$ for $\mathscr{IG}(1,1)$. \label{tab:IG}}
		\vspace{-0.25cm}
	\end{center}
\end{table}


\subsubsection{The L\'evy Distribution}

We last consider the L\'evy density, $f_Z(z) \mspace{-3mu} \sim \mspace{-3mu} \mathscr{L}(\mu,b)$, with a {\em location} parameter\footnote{Note that $\mu$ is {\em not} the mean, as the mean of the L\'evy density is $\infty$.} $\mu$ and a scale parameter $b \mspace{-3mu} > \mspace{-3mu} 0$.
Similarly to the IG density, in Appendix \ref{annex:CondsLemmaFA} it is shown that for $f_Z(z) \mspace{-3mu} \sim \mspace{-3mu} \mathscr{L}(\mu,b)$, the conditions of Lemma \ref{lemm:unimodalFA} hold and therefore $f_Z(z)$ is unimodal for sufficiently large $M$. Moreover, explicitly evaluating $\sf{D}_{\text{ML}}$ seems intractable, yet, it can be evaluated numerically. For the FA detector we use the CDF of the L\'evy density to obtain $\sf{D}_{\text{FA}} \mspace{-3mu} = \mspace{-3mu} - \log \left(1 \mspace{-3mu} - \mspace{-3mu} \text{erfc}\left(\sqrt{\frac{b}{2\Delta}} \right) \right)$, where $\text{erfc}(\cdot)$ is the complementary error function. As for the linear detector we recall Remark \ref{rem:destructiveLinear} which states that for L\'evy-based propagation $\sf{D}_{\text{LIN}} \mspace{-3mu} = \mspace{-3mu} 0$.

Table \ref{tab:Levy} details the system diversity gain for $\mu \mspace{-3mu} = \mspace{-3mu} 0,b \mspace{-3mu} = \mspace{-3mu} 1$, and $\Delta \mspace{-3mu} \in \mspace{-3mu} \{0.5, 1, 1.5 \}$.
\begin{table}[h]
		\begin{center}
		\footnotesize
		\begin{tabular}[t]{|c|c|c|c|}
			\hline
			  & $\Delta=0.5$  & $\Delta=1$ & $\Delta=1.5$ \\
			\hline
			\hline
			$\sf{D}_{\text{ML}}$  &  0.1791 &  0.3828 &  0.5350 \\
			\hline
			$\sf{D}_{\text{FA}}$  & 0.1711  & 0.3817  &  0.5348  \\
			\hline 
		\end{tabular}
		\captionsetup{font=footnotesize}
		\caption{$\sf{D}_{\text{ML}}$ and $\sf{D}_{\text{FA}}$ for $\mathscr{L}$(0,1). \label{tab:Levy}}
		\vspace{-0.25cm}
	\end{center}
\end{table}

It can be clearly observed that the gap in system diversity gain between the FA and ML detectors is very small for both the IG and L\'evy densities. 

Next, we consider noise densities with a finite support (i.e., $\tau < \infty$).

\subsection{Densities With $\tau < \infty$}

\begin{figure*}[t]
	\normalsize
	\captionsetup{font=footnotesize}
	\centering
	\begin{minipage}{.475\textwidth}
		\centering
		\includegraphics[width=.8\linewidth]{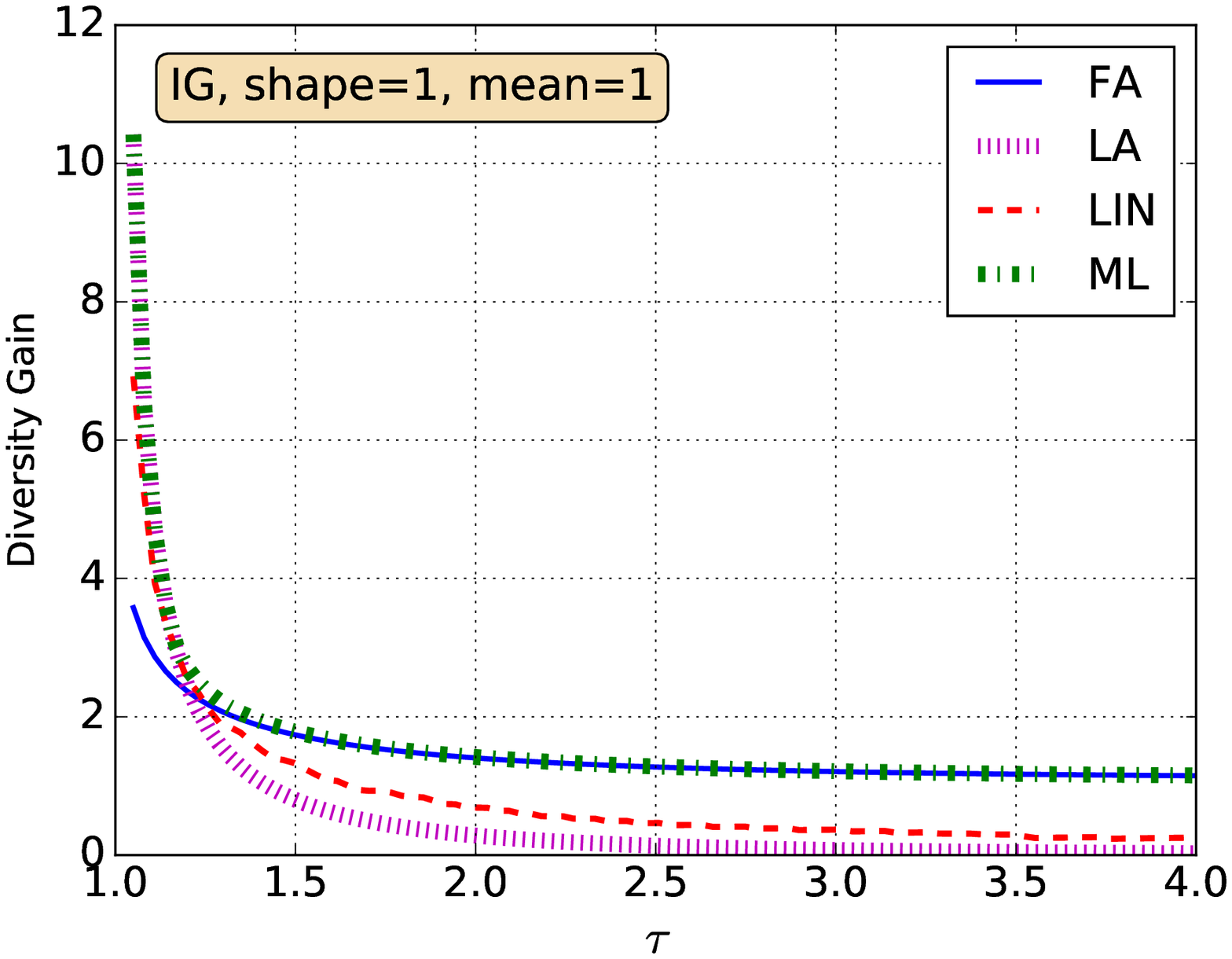}
		\captionof{figure}{Diversity gain versus $\tau$ for $Z \sim \mathscr{IG}(1,1,\tau)$ and $\Delta = 1$.}
		\label{fig:IG}
	\end{minipage}
	\hspace{0.6cm}
	\begin{minipage}{.475\textwidth}
		\centering
		\includegraphics[width=.8\linewidth]{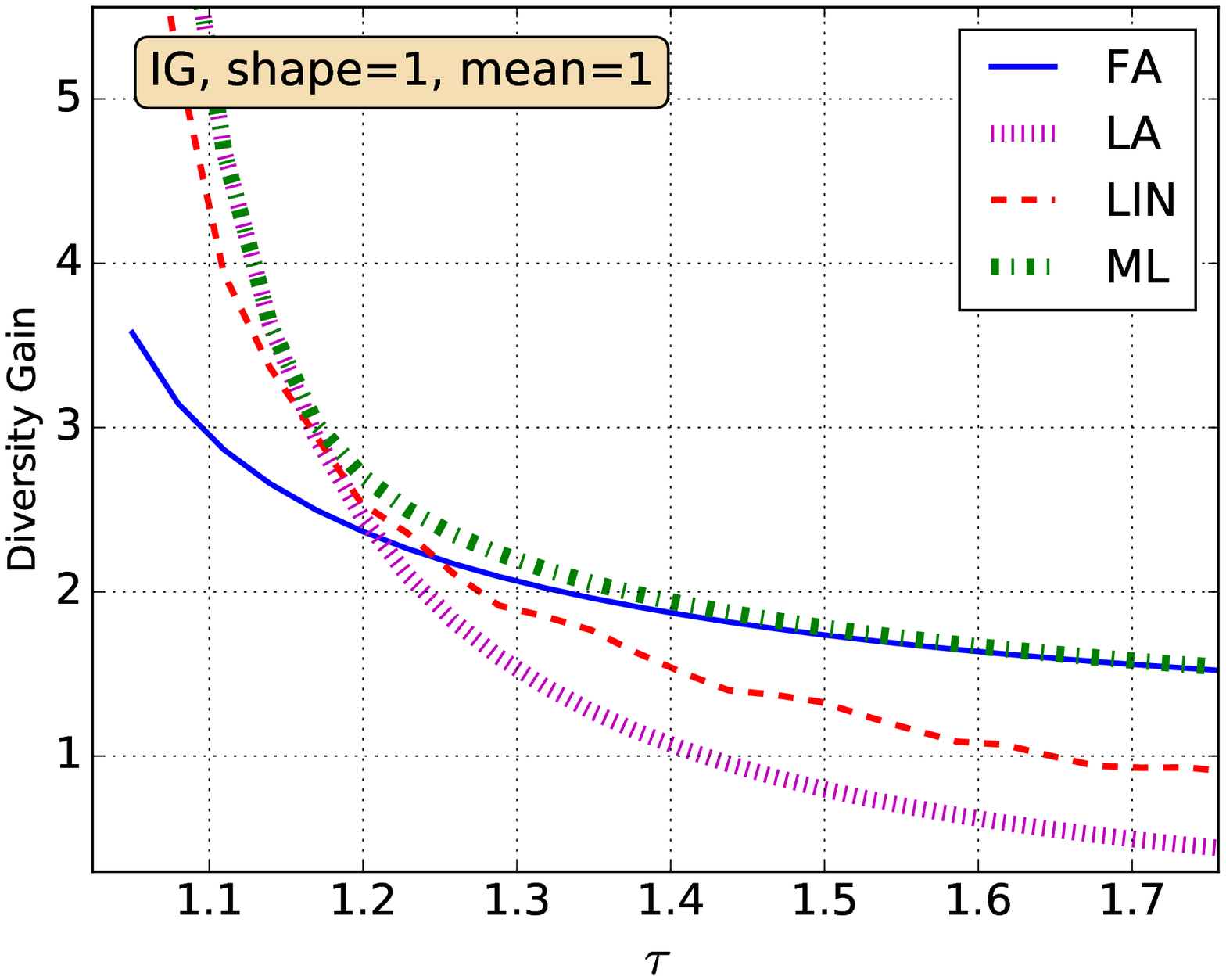}
		\captionof{figure}{Diversity gain versus $\tau$ for $Z \sim \mathscr{IG}(1,1,\tau)$ and $\Delta = 1$ (zoomed).}
		\label{fig:IG_zoomed}
	\end{minipage}
	\begin{minipage}{.475\textwidth}
		\centering
		\includegraphics[width=.8\linewidth]{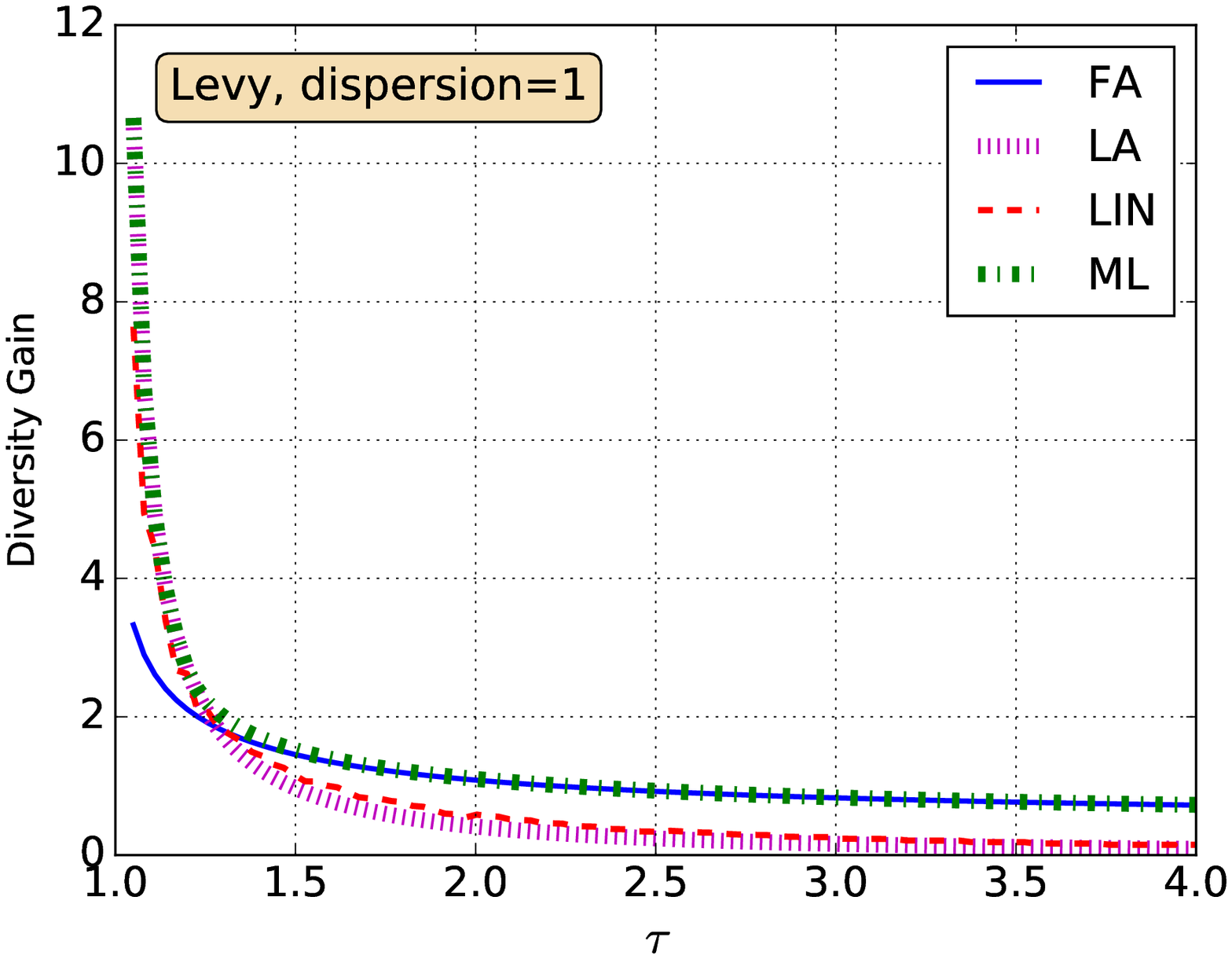}
		\captionof{figure}{Diversity gain versus $\tau$ for $Z \sim \mathscr{L}(0,1,\tau)$ and $\Delta = 1$.}
		\label{fig:Levy}
	\end{minipage}
	\hspace{0.6cm}
	\begin{minipage}{.475\textwidth}
		\centering
		\includegraphics[width=.8\linewidth]{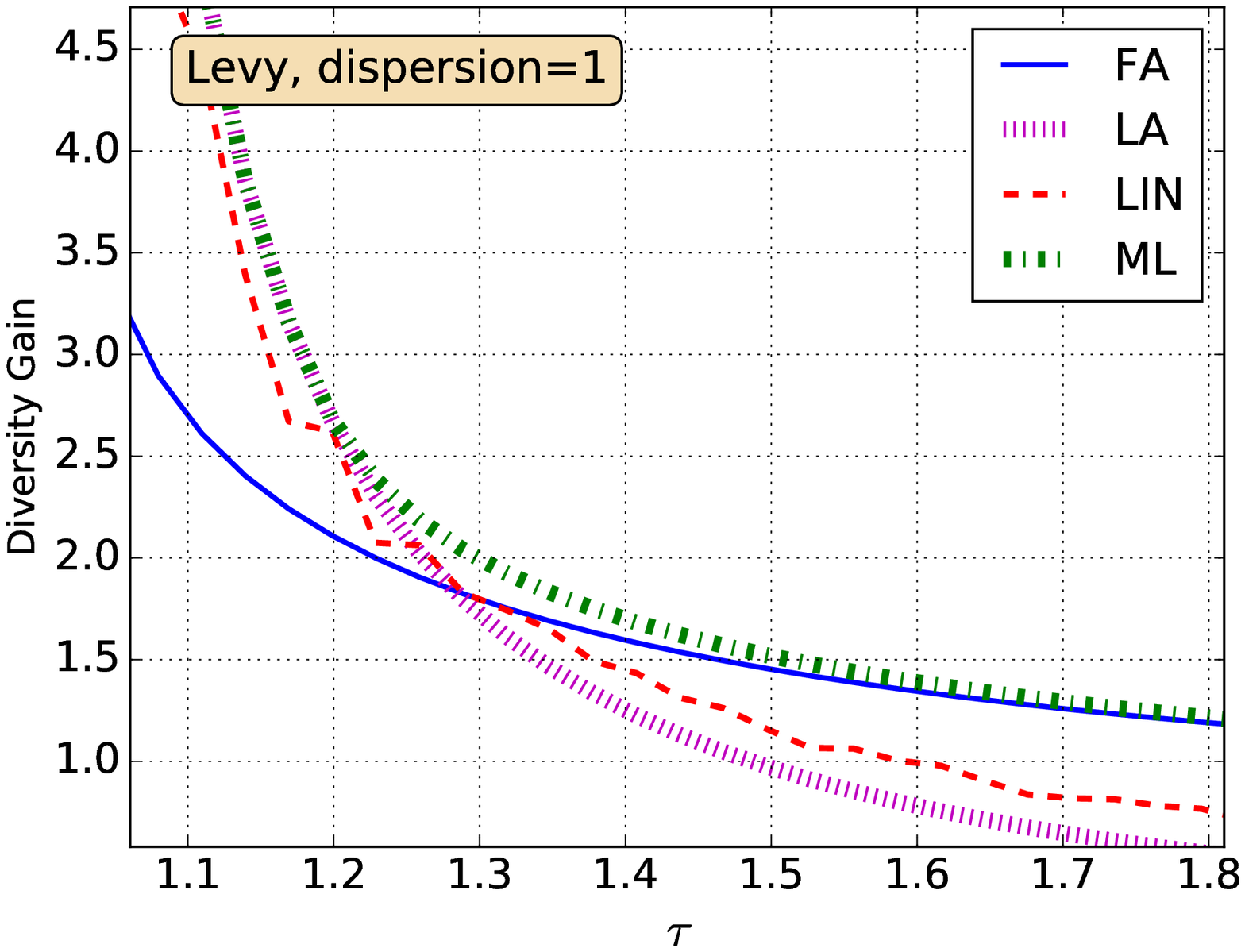}
		\captionof{figure}{Diversity gain versus $\tau$ for $Z \sim \mathscr{L}(0,1,\tau)$ and $\Delta = 1$ (zoomed).}
		\label{fig:Levy_zoomed}
	\end{minipage}
	\end{figure*}

We first consider the uniform distribution supported over $[0, \tau]$. Since the uniform distribution is constant over its support, according to Def. \ref{def:unimodal} the whole support constitutes the mode of the density (in particular, both $z=0$ and $z = \tau$). 

\subsubsection{The Uniform Distribution} \label{subsec:unifom}

Let $f_Z(z) \mspace{-3mu} \sim \mspace{-3mu} \mathscr{U}(0,\tau), \tau \mspace{-3mu} > \mspace{-3mu} \Delta$, i.e., the uniform density over $[0,\tau]$. 
Following Corollaries \ref{cor:zreomode} and \ref{cor:taumode}, $\sf{D}_{\text{ML}} \mspace{-1mu} = \mspace{-1mu} \sf{D}_{\text{FA}} \mspace{-1mu} = \mspace{-1mu} \sf{D}_{\text{LA}} \mspace{-1mu} = \mspace{-1mu} \log \frac{\tau}{\tau-\Delta}$.
For the linear detector we do not have a closed form expression for $\Lambda_Z^{\ast}(v)$. However, we can use the moment generating function (MGF) of the uniform distribution and numerically find $\sf{D_{\text{LIN}}}$.

For instance, let $\tau \mspace{-3mu} = \mspace{-3mu} 1$, and consider $\Delta \mspace{-3mu} \in \mspace{-3mu} \{ 0.25, 0.5, 0.75 \}$. Table \ref{tab:uniform} details the resulting diversity gains. The table indicates large performance gains of the FA and LA detectors over linear detection.
\begin{table}[h]
		\begin{center}
		\footnotesize
		\begin{tabular}[t]{|c|c|c|c|}
			\hline
			  & $\Delta=0.25$  & $\Delta=0.5$ & $\Delta=0.75$ \\
			\hline
			\hline
			$\sf{D}_{\text{ML}}, \sf{D}_{\text{FA}}, \sf{D}_{\text{LA}}$  & 0.2879  & 0.6931  & 1.3863  \\
			\hline
			$ \sf{D}_{\text{LIN}}$  & 0.0956  & 0.4086  & 1.0798 \\
			\hline 
		\end{tabular}
		\captionsetup{font=footnotesize}
		\caption{$\sf{D}_{\text{ML}}, \sf{D}_{\text{FA}}$ and $\sf{D}_{\text{LIN}}$ for $\mathscr{U}(0,1)$. \label{tab:uniform}}
		\vspace{-0.25cm}
	\end{center}
\end{table}

Next, we consider the truncated (clipped) L\'evy and IG distributions.

\subsubsection{The Truncated L\'evy and IG Distributions} 
Considering the truncated L\'evy and IG distributions is motivated by scenarios where the information particles degrade over time \cite{ITsubmission, Pandey17}. Specifically, these truncated distributions model a {\em finite lifespan} of the particles, where the particles are dissipated immediately after the time interval $[0,\tau]$. Such truncated noise densities can be (approximately) achieved by using enzymes or other chemicals to quickly degrade the particles \cite{noe14, narimanSpawc2016, guo15}.  

Following the approach taken in Section \ref{subsec:compare_FALA}, we let $\tilde{f}_Z(z)$ be a continuous, differentiable, and unimodal density with support $\realSet^{+}$. The noise density $f_Z(z)$ is obtained by truncating $\tilde{f}_Z(z)$ at $\tau$. As stated in Section \ref{subsec:compare_FALA}, the PDF and CDF are given by $f_Z(z) = \frac{\tilde{f}_Z(z)}{\tilde{F}_Z(\tau)}$ and $F_Z(z) = \frac{\tilde{F}_Z(z)}{\tilde{F}_Z(\tau)}$, respectively.

Recall that the L\'evy or IG densities correspond to diffusion with and without a drift, respectively, and let $\mathscr{IG}(\mu,b,\tau)$ denote the IG distribution truncated at $\tau$. Similarly, let $\mathscr{L}(\mu,b,\tau)$ denote the L\'evy distribution truncated at $\tau$. Figs. \ref{fig:IG} and \ref{fig:IG_zoomed} depict the diversity gain achieved by the ML, linear, FA, and LA detectors, for $Z \sim \mathscr{IG}(1,1,\tau)$. It can be observed that for large values of $\tau$ the FA detector achieves diversity gain very close to the ML, while the LA and linear detectors perform poorly. This extends the results reported in Table \ref{tab:IG}. On the other hand, when $\tau$ is small the diversity gain achieved by the LA detector is very close to that achieved by the ML detector; in fact, the curves are practically indistinguishable. In this regime the FA detector performs poorly, while the linear detector is superior to the FA yet inferior to the LA. Using \eqref{eq:FA_eq_LA} we have that the curves corresponding to the FA and LA detector intersect at $\tau^{\ast} = 1.208$. Indeed, Fig. \ref{fig:IG} indicate that by using the LA detector for $\tau < 1.208$ and the FA detector for $\tau \ge 1.208$, one achieves diversity gain very close to the diversity gain achieved by the ML detector.

Figs. \ref{fig:Levy} and \ref{fig:Levy_zoomed} depict the diversity gain achieved by the ML, linear, FA, and LA detectors, for $Z \sim \mathscr{L}(0,1,\tau)$. Similarly to the truncated IG, the FA performs well for large values of $\tau$ (which supports the results reported in Table \ref{tab:Levy}), while the LA performs well for small values of $\tau$. Here, \eqref{eq:FA_eq_LA} leads to $\tau^{\ast} = 1.282$. 

\vspace{-0.15cm}
\section{Conclusion} \label{sec:conc}

We have studied one-shot communication over \ymc{molecular} timing channels assuming that $M$ information particles are simultaneously released and that their propagation follows a unimodal density. 
We defined the system diversity gain $\sf{D}$ to be the {\em exponential} rate of decrease of the detection probability of error $P_{\varepsilon}^{(M)}$ when $M$ grows asymptotically large. We then derived closed form expressions for the $\sf{D}$ achievable by four detectors: the optimal ML detector, a linear detector, the FA detector, and the LA detector. 
We showed that the FA detector achieves a diversity gain very close to that of the ML detector, while being simpler and having significantly shorter delays, when the density of the noise is supported over a large interval (for instance $\realSet^+$). In particular, for delay densities where the mode of the density is zero, the FA detector is optimal. 
We also showed that when the density of the noise is supported over a small interval, the LA detector achieves a diversity gain very close to that of the ML detector. Particularly, for delay densities where the mode of the density is at the maximum of its support, the LA detector is optimal. 
Our numerical evaluations show that by combining the FA and LA detectors one can achieve performance very close to that of the ML detector for all ranges of support intervals. Specifically, for almost every support interval, this combined detector outperforms the linear detector. We conclude that this combined detector constitutes a low-complexity near-ML detection framework for one-shot communication over timing channels. 

\appendices

\numberwithin{equation}{section}

\section{Proof of Lemma \ref{lemm:unimodalFA}} \label{annex:proof_lemm_unimodalFA}

A unimodal density supported on $\realSet^{+}$ belongs to one of the following classes of densities: 
\begin{enumerate}
	\item
		Unimodal densities with mode $m_Z \mspace{-3mu} = \mspace{-3mu} 0$.
		
	\item
		Unimodal densities with $m_Z \mspace{-3mu} > \mspace{-3mu} 0$, and $\lim_{z \to 0^+} f_Z(z) \mspace{-3mu} = \mspace{-3mu} \nu \mspace{-3mu} > \mspace{-3mu} 0$.
		
	\item
		Unimodal densities with $m_Z \mspace{-3mu} > \mspace{-3mu} 0$, and $\lim_{z \to 0^+} f_Z(z) \mspace{-3mu} = \mspace{-3mu} 0$.
	
\end{enumerate}
We next show that densities from the first two classes are unimodal for sufficiently large $M$ regardless of the conditions of Lemma \ref{lemm:unimodalFA}. Then, we show that the conditions stated in Lemma \ref{lemm:unimodalFA} ensure that densities from the third class are unimodal for sufficiently large $M$. 

From Def. \ref{def:unimodal}, if a unimodal density has more than a single maximum, then the maximum must be a continuous interval. Thus, the derivative of the density changes its sign at most once. Next, recall that $f_{Z_{\mathrm{FA}}}(z) \mspace{-3mu} = \mspace{-3mu} M \mspace{-3mu} \cdot \mspace{-3mu} f_Z(z) \mspace{-3mu} \cdot \mspace{-3mu} (1 \mspace{-3mu} - \mspace{-3mu} F_Z(z))^{M-1}$. Hence, the derivative of $f_{Z_{\mathrm{FA}}}(z)$ is given by:
\begin{align}
	f^{'}_{Z_{\mathrm{FA}}}(z) \mspace{-3mu} & = \mspace{-3mu} M \left( f^{'}_Z(z)(1 \mspace{-3mu} - \mspace{-3mu} F_Z(z))^{M-1} \mspace{-3mu} \right. \nonumber \\
	& \mspace{70mu} \left. - f^2_Z(z)(M-1)(1 \mspace{-3mu} - \mspace{-3mu} F_Z(z))^{M-2} \right).
\end{align}

Setting $f^{'}_{Z_{\mathrm{FA}}}(z) \mspace{-3mu} = \mspace{-3mu} 0$ we obtain conditions indicating when $f_{Z_{\mathrm{FA}}}(z)$ decreases:
\begin{align}
	f^{'}_{Z_{\mathrm{FA}}}(z) \mspace{-3mu} \le \mspace{-3mu} 0 \quad \Leftrightarrow \quad \frac{f^{'}_Z(z)(1 \mspace{-3mu} - \mspace{-3mu} F_Z(z))}{f^2_Z(z)} \mspace{-3mu} \le \mspace{-3mu} M-1. \label{eq:deriv_negative}
\end{align} 

\noindent For densities that belong to the first class we have $f^{'}_Z(z) \mspace{-3mu} \le \mspace{-3mu} 0$. Therefore, as $(1 \mspace{-3mu} - \mspace{-3mu} F_Z(z))$ and $f^2_Z(z)$ are positive, $f^{'}_{Z_{\mathrm{FA}}}(z)$ is non-increasing and unimodal for {\em any} $M$. 

For the second class we note that in the range $0 \mspace{-3mu} < \mspace{-3mu} z \mspace{-3mu} \le \mspace{-3mu} m_Z$, $f^2_Z(z) \mspace{-3mu} \ge \mspace{-3mu} \nu^2$. Thus, $\frac{f^{'}_Z(z)(1 \mspace{-3mu} - \mspace{-3mu} F_Z(z))}{f^2_Z(z)}$ is positive and bounded, and by choosing $M$ large enough $f_{Z_{\mathrm{FA}}}(z)$ is decreasing for any $z$ and therefore unimodal. 

Finally, for densities in the third class with $\lim_{z \to 0^+} f_Z(z) \mspace{-3mu} = \mspace{-3mu} 0$, since $f_Z(z)$ is assumed to be differentiable, it is possible that:\footnote{Recall that since $m_Z \mspace{-3mu} > \mspace{-3mu} 0$, then there exists an $\epsilon>0$ such that $0 \mspace{-3mu} \le \mspace{-3mu}f^{'}_Z(z), 0 \mspace{-3mu} < \mspace{-3mu} z \mspace{-3mu} < \mspace{-3mu} \epsilon$.}
\begin{align}
	\lim_{z \to 0^+} \frac{f^{'}_Z(z)(1 \mspace{-3mu} - \mspace{-3mu} F_Z(z))}{f^2_Z(z)} \mspace{-3mu} = \mspace{-3mu} \infty.
\end{align}

\noindent Since $(1 \mspace{-3mu} - \mspace{-3mu} F_Z(z))$ is monotonically decreasing with $z$, requiring that $\frac{f^{'}_Z(z)}{f^2_Z(z)}$ will decrease monotonically for $0 \mspace{-3mu} < \mspace{-3mu} z \mspace{-3mu} < \mspace{-3mu} \epsilon$ ensures that $\frac{f^{'}_Z(z)(1 \mspace{-3mu} - \mspace{-3mu} F_Z(z))}{f^2_Z(z)}$ will also be monotonically decreasing. In such case there is a $z_0$ for which:
\begin{align}
	\frac{f^{'}_Z(z)(1 \mspace{-3mu} - \mspace{-3mu} F_Z(z))}{f^2_Z(z)} \begin{matrix} z \mspace{-3mu} < \mspace{-3mu} z_0 \\ \gtrless \\ z \mspace{-3mu} > \mspace{-3mu} z_0 \end{matrix} M-1.
\end{align}

Hence, the density is unimodal for all $M \mspace{-3mu} > \mspace{-3mu} M_0$, where $M_0$ is given by $M_0 \mspace{-4mu} = \mspace{-4mu} \left\lceil \mspace{-3mu} \frac{f^{'}_Z(\xi)(1 \mspace{-3mu} - \mspace{-3mu} F_Z(\xi))}{f^2_Z(\xi)} \right\rceil \mspace{-4mu} + \mspace{-4mu} 1$
%
\noindent and $\xi \mspace{-4mu} = \mspace{-3mu} \argmax_z \frac{f^{'}_Z(z)(1 \mspace{-3mu} - \mspace{-3mu} F_Z(z))}{f^2_Z(z)}, z \mspace{-3mu} > \mspace{-3mu} \epsilon$.
 

\section{Proof of Lemma \ref{lemm:unimodalLA}} \label{annex:proof_lemm_unimodalLA}

Similarlly to the derivation in Appendix \ref{annex:proof_lemm_unimodalFA},  Def. \ref{def:unimodal} implies that if a unimodal density has more than a single maximum, then the maximum must be a continuous interval. Thus, the derivative of the density changes its sign at most once. Next, recall that $f_{Z_{\mathrm{LA}}}(z) \mspace{-3mu} = \mspace{-3mu} M \mspace{-3mu} \cdot \mspace{-3mu} f_Z(z) \mspace{-3mu} \cdot \mspace{-3mu} (F_Z(z))^{M-1}$. Hence, the derivative of $f_{Z_{\mathrm{LA}}}(z)$ is given by:
\begin{align}
	\mspace{-8mu} f^{'}_{Z_{\mathrm{LA}}}(z) \mspace{-4mu} & = \mspace{-4mu} M F_Z(z)^{M-2} \left( \mspace{-3mu} f^{'}_Z(z) F_Z(z) \mspace{-3mu} + \mspace{-3mu} f^2_Z(z)(M \mspace{-4mu}- \mspace{-4mu} 1) \mspace{-3mu} \right) \mspace{-3mu} .
\end{align}

Setting $f^{'}_{Z_{\mathrm{LA}}}(z) \mspace{-3mu} = \mspace{-3mu} 0$ we obtain conditions indicating when $f_{Z_{\mathrm{LA}}}(z)$ decreases:
\vspace{-0.15cm}
\begin{align}
	f^{'}_{Z_{\mathrm{LA}}}(z) \mspace{-3mu} \le \mspace{-3mu} 0 \quad \Leftrightarrow \quad \frac{f^{'}_Z(z) F_Z(z)}{f^2_Z(z)} \mspace{-3mu} \le \mspace{-3mu} -(M-1). \label{eq:deriv_negative_LA}
\end{align} 

\noindent Thus we need to show that {\em if} $f_{Z_{\mathrm{LA}}}(z)$ {\em starts decreasing it does not increase again}. To show this we recall that the support of $f_Z(z)$ is $(0,\tau)$ and that $|f^{'}_Z(z)| < \infty$. We now consider two cases: 
\begin{enumerate}
	\item 
		$\lim_{z \to \tau} f_Z(z) = \nu > 0$
		
	\item 
		$\lim_{z \to \tau} f_Z(z) = 0$		
\end{enumerate}

 For the first case \eqref{eq:deriv_negative_LA} implies that then there exist an $M_0$ such that for $M>M_0$, $f_{Z_{\mathrm{LA}}}(z)$ is increasing for all $z$ in $(0,\tau)$. Thus, in this case $f_{Z_{\mathrm{LA}}}(z)$ is clearly unimodal.

For the second case we note that \eqref{eq:deriv_negative_LA} holds only if $f^{'}_Z(z) < 0$ (since $F_Z(z)$ and $f^2_Z(z)$ are positive). 
Moreover, only when $z \to \tau$ it is possible that: 
\begin{align}
	\lim_{z \to \tau} \frac{f^{'}_Z(z) F_Z(z)}{f^2_Z(z)} = -\infty, 
\end{align}

\noindent The condition of the lemma ensures that there is an interval $(\tau-\epsilon, \tau), \epsilon > 0$, where $\frac{f^{'}_Z(z) F_Z(z)}{f^2_Z(z)}$ {\em monotonically} decreases. Therefore, by choosing $M$ large enough it can be guaranteed that once $f_{Z_{\mathrm{LA}}}(z)$ starts decreasing it does not increase again, thus, it is unimodal.

\section{The Conditions of Lemma \ref{lemm:unimodalFA} and Lemma \ref{lemm:unimodalLA} for the L\'evy and IG Densities} \label{annex:CondsLemmaFA}

In this section we evaluate the function $g(z) = \frac{f^{'}_Z(z)}{f^2_Z(z)}$ for the L\'evy and IG Densities. We show that for $z$ small enough the derivative of $g(z)$ is negative and therefore it monotonically decreases as required.
We begin with the L\'evy distribution where we write:
\begin{align}
	g_{\mathrm{Lev}}(z) = \frac{f'_Z(z)}{f^2_Z(z)} = \sqrt{\frac{\pi}{2 c z}} e^{\frac{c}{2z}} (c - 3z). \label{eq:lev_g}
\end{align}

\noindent Thus, $g'(z)$ is given by:
\begin{align}
	g'_{\mathrm{Lev}}(z) = - \sqrt{\frac{\pi}{8 c z^5}} e^{\frac{c}{2z}} (c^2 - 2cz + 3z^2). \label{eq:lev_g_tag}
\end{align}

\noindent Observe that $g'_{\mathrm{Lev}}(z) < 0$ for any finite $z$ and $c > 0$. Thus, for the L\'evy distribution, the conditions of both lemmas hold.

Next, we consider the IG distribution, where we have:
\begin{align}
	g_{\mathrm{IG}}(z) = \sqrt{\frac{\pi}{2 \mu^4 b z}} e^{\frac{b (z - \mu)^2}{2 \mu^2 z}} (b (\mu^2 - z^2) - 3\mu^2 z), 	\label{eq:IG_g}
\end{align}

\noindent and
\begin{align}
	g'_{\mathrm{IG}}(z) & = - \sqrt{\frac{\pi}{8 \mu^8 b z^{5}}} e^{\frac{b(z-\mu)^2}{2 \mu^2 z}} \nonumber \\
	& \mspace{30mu} \times (3 \mu^4 z^2 \mspace{-3mu} + \mspace{-3mu} b^2 (\mu^2 \mspace{-3mu} - \mspace{-3mu} z^2)^2 \mspace{-3mu} + \mspace{-3mu} b (6 \mu^2 z^3 \mspace{-3mu} - \mspace{-3mu} 2 \mu^4 z)).
\end{align}

\noindent Again, it can be shown that $g'_{\mathrm{IG}}(z) < 0$ for any finite $z>0$ and $\mu,b > 0$. Thus, for the IG distribution, the conditions of both lemmas hold.

\section*{Acknowledgment}

The authors would like to thank Andrea Montanari for comments which greatly simplified the proof of Thm. \ref{thm:FA_PDG}.



\end{document}